%% file: main.tex
\documentclass{sig-alternate}
\newcommand{\ignore}[1]{}
\usepackage[pass]{geometry}
\usepackage{fancyhdr}
\usepackage[normalem]{ulem}
\usepackage[hyphens]{url}
\usepackage{hyperref}
\usepackage{color}
\usepackage{soul}

\usepackage{enumitem}
\usepackage{tikz}
\newcommand{\ballnumber}[1]{\tikz[baseline=(myanchor.base)] \node[circle,scale=0.9, fill=.,inner sep=1pt] (myanchor) {\color{-.}\bfseries\footnotesize #1};}

\makeatletter
\newcommand\subparagraph{%
  \@startsection{subparagraph}{5}
  {\parindent}
  {3.25ex \@plus 1ex \@minus .2ex}
  {-1em}
  {\normalfont\normalsize\bfseries}}
\makeatother
\usepackage{titlesec}
\let\subparagraph\relax 

\begin{document}
\title{AMOEBA: A Coarse Grained Reconfigurable Architecture for Dynamic GPU Scaling}








\author{%
Xianwei Cheng\\
 \affaddr{Computer Science and Engineering Department}\\
 \affaddr{University of North Texas}\\
 \email{xianweicheng@my.unt.edu}
\and
Hui Zhao\\
 \affaddr{Computer Science and Engineering Department}\\
 \affaddr{University of North Texas}\\
 \email{hui.zhao@unt.edu}
\and
Mahmut Kandemir\\
 \affaddr{Computer Science and Engineering Department}\\
 \affaddr{Pennsylvania State University}\\
 \email{mtk2@psu.edu}
\and
Beilei Jiang\\
 \affaddr{Computer Science and Engineering Department}\\
 \affaddr{University of North Texas}\\
 \email{beileijiang@my.unt.edu}
\and
Gayatri Mehta\\
 \affaddr{Electrical Engineering Department}\\
 \affaddr{University of North Texas}\\
 \email{gayatri.mehta@unt.edu}
\and
}
\maketitle


\maketitle
\pagestyle{plain}

\titlespacing*{\section} {0pt}{2ex plus 0ex minus 0ex}{1ex plus 0ex}

\titlespacing*{\subsection} {0pt}{1.6ex plus 0ex minus 0ex}{0.6ex plus 0ex}

\titlespacing*{\subsubsection} {0pt}{1.2ex plus 0ex minus 0ex}{0.2ex plus 0ex}


\begin{abstract}

Different GPU applications exhibit varying scalability patterns with network-on-chip (NoC), coalescing, memory and control divergence, and L1 cache behavior. A GPU consists of several Streaming Multi-processors (SMs) that collectively determine how shared resources are partitioned and accessed. Recent years have seen divergent paths in SM scaling towards scale-up (fewer, larger SMs) vs. scale-out (more, smaller SMs). 
However, neither scaling up nor scaling out can meet the scalability requirement of all applications running on a given GPU system, which inevitably results in performance degradation and resource under-utilization for some applications. In this work, we investigate major design parameters that influence GPU scaling. We then propose AMOEBA, a solution to GPU scaling through reconfigurable SM cores.  AMOEBA monitors and predicts application scalability at run-time and adjusts the SM configuration to meet program requirements. AMOEBA also enables dynamic creation of heterogeneous SMs through independent fusing or splitting. AMOEBA is a microarchitecture-based solution and requires no additional programming effort or custom compiler support.
Our experimental evaluations with application programs from various benchmark suites indicate that AMOEBA is able to achieve a maximum performance gain of 4.3x, and generates an average performance improvement of 47\% when considering all benchmarks tested. 

\end{abstract}

\input{introduction}

\input{design}
\input{evaluation}

\input{relatedwork}

\bibliographystyle{ieeetr}
\bibliography{ref}

\end{document}

%% file: introduction.tex

\section{Introduction}

GPUs have emerged as performance accelerators for general purpose applications and take advantage of the single instruction multiple threads (SIMT) programming model to improve the performance of data-parallel computations. Supercomputers \cite{super1, super2}, cloud servers \cite{web}, desktops \cite{desktop} and even mobile devices \cite{mobile} already benefit significantly from GPUs to achieve better performance and higher power efficiency. A GPU typically consists of many compute units (CUs), also called streaming multiprocessors (SMs), and each SM contains a large number of simple compute cores \cite{SM}. GPUs leverage the massive number of computing cores in SMs to exploit thread-level parallelism (TLP) in an attempt to hide memory access latency \cite{mem1, mem2}. 

The multiprocessor industry has been fueled by Moore's Law for many years and processor performance is improved through increasing transistor counts. However, Moore's Law  is slowing down because we are reaching the technological limits of how small transistors can be made. Increasing the chip size can allow us to add more transistors to a chip. However, this is not a sustainable solution due to several reasons: (1) there is not enough power budget to allow all transistors to be powered on simultaneously. This is because transistor threshold voltage does not scale with technology nodes and per-transistor switching energy almost keeps constant \cite{compositecores}, (2) cost becomes prohibitive in manufacturing chips with ultra-high transistor counts \cite{cost}, and (3) data communication becomes a bottleneck as chip size increases \cite{checkerboard}. 

TLP has been considered as a promising solution to tackle the slowdown of Moore's law, and GPU architecture is based on the idea of exploiting TLP. The computing power of GPU arises from its SIMT architecture: many threads are executed concurrently in an SIMD fashion. However, an average programmer may not be aware of the details of the underlying hardware to write high-quality code to fully utilize available GPU resources. As a result, many general purpose applications are not fully optimized for running on a specific GPU architecture, and this causes under-utilization of hardware resources \cite{SMK, SGMF, rgpu, datapathfusion, GPUMaestro, multiprogram, memorystall}. For example, it has been observed that cores are idle for 52\% to 98\% of the execution time for some GPU benchmarks \cite{SMK}. Therefore, instead of adding more resources to GPUs, exploration of optimized resource utilization techniques can be a more viable option to enhance the GPU performance and power efficiency.

There have been earlier efforts targeting to maximize GPU resource utilization. For example, several prior works have proposed to share a GPU among multiple applications and used software-level techniques to manage resource sharing \cite{software1, software2, software3}. On the hardware side, spatial multitasking has been proposed as a technique that partitions a GPU among multiple kernels at an SM granularity \cite{SMT}. Several techniques have been also proposed to share resources among kernels inside each SM, such as simultaneous multikernel (SMK) \cite{SMK}, warp-slicer \cite{warpslicer}, and GPU Maestro \cite{GPUMaestro}. However, these techniques need to run multi-programmed workloads to fully utilize GPU resources (i.e., finding more tasks to avoid GPU inactive cycles). Also, they do not consider an application's scalability and do not configure hardware to meet the software's resource demands.

An alternative approach is to dynamically \textit{reconfigure} hardware to avoid resource under-utilization. Reconfigurable cores have been proposed in the past for CPUs to facilitate parallel execution \cite{corefusion, compositecores, voltron, trips, smartmemory}. However, the overhead of reconfiguring CPU cores is high due to the complexities associated with CPU architectures.  There have been very few reconfigurable GPU architectures proposed. For example, R-GPU is proposed to interconnect compute cores inside an SM to reduce data movement and remove decoding overhead by assigning each core a fixed operation \cite{rgpu}. In comparison, SGMF is proposed  as a  dataflow  architecture  using  coarse-grain reconfigurable fabric \cite{SGMF}. However, SGMF needs help from compiler to convert the  kernels into dataflow graphs. Neither work considers the scalability of applications regarding system bottlenecks, such as interconnection network, memory access patterns, and control divergence. In addition, these prior works only explore intra-SM resource utilization and assume that the number and size of SMs are fixed. However, sharing resources among SMs is also important because applications have varying scalability patterns depending on SM settings, but exploration of this design space has largely been ignored in prior work.

In this work, we present AMOEBA, a reconfigurable architecture to improve GPU resource utilization, performance, and energy efficiency. AMOEBA takes into account several important application resources requirements such as interconnect throughput, memory access patterns, and control divergence, before selecting an optimal GPU scaling option. AMOEBA is a coarse-grained reconfigurable architecture to enable flexible SM scaling at a low cost. It also explores heterogeneity of SMs through dynamic fusing and splitting in order to accommodate program divergence.

We make the following {\bf contributions} in this paper:
\begin{itemize}[topsep=0pt,parsep=0pt,partopsep=0pt,leftmargin=10pt,labelwidth=6pt,labelsep=4pt]
\item We investigate the GPU \textit{scaling} problem under resource bound. We identify the important factors determining whether an SM should be designed in a \textit{scale-up} or \textit{scale-out} fashion. Building upon the results from this investigation, we propose a coarse-grained reconfigurable architecture that fuses the baseline scale-out SMs into larger scale-up SMs. This design enables optimized resource utilization across SM boundaries.
\item We design an online \textit{controller} that takes into account an application's dynamic behavior and makes reconfiguration decisions accordingly. The controller employs a \textit{binary logistic regression model} to predict application scalability with low cost.
\item We provide design details of the proposed reconfigurable architecture. Our proposed architecture enables coarse-grained SM fusion, and it can provide support for both scale-up and scale-out GPU configurations.
\item We propose a scheme to split individual scale-up SM cores dynamically when program divergence causes pipeline stalls.
\end{itemize}
\vspace{-0.02in}
To the best of our knowledge, this is the first paper to propose a reconfigurable GPU architecture that can dynamically toggle between scale-up and scale-out options, with the goal of maximizing resource utilization.








\begin{figure}[t]
  \centering
  \scalebox{0.9}{\epsfig{file=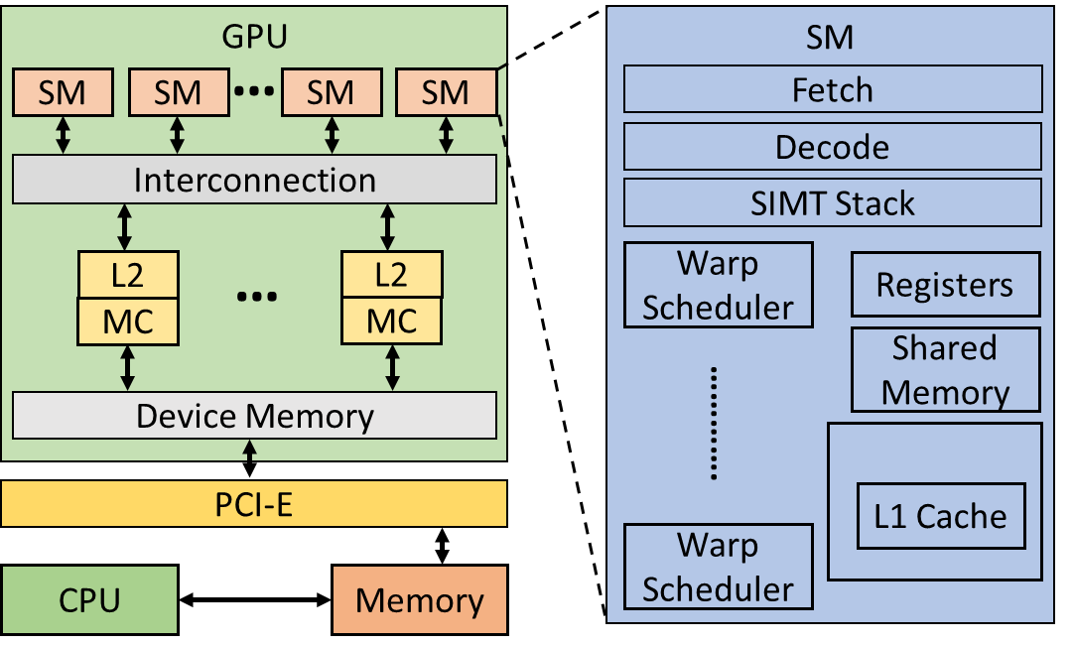,width=0.4\textwidth,clip=}}
   \vspace{-0.23cm}
 \caption{ GPU architecture overview.} 
   \label{fig:system}
\vspace{-0.18 in}
\end{figure} 
 \section{Background}
\subsection{GPU Execution Model and Architecture}
GPU execution model divides the total work space into a grid and assigns a work item, also called a thread, to work on each portion of data. Each thread executes the same set of instructions, and this enables parallel multi-threaded execution in an SIMT fashion. Each segment of code loaded to a GPU is called a kernel. A group of threads that execute a kernel concurrently is referred to as a workgroup, also called a thread-block or a cooperative thread-array (CTA). The total work space is divided into blocks or CTAs, and threads within a given CTA can communicate with each other.

A high-level view of a GPU architecture is shown in Figure~\ref{fig:system}. A GPU consists of multiple compute units (CUs) or streaming multiprocessors (SMs), which are analogous to CPU cores. Each SM contains fetch, decode, execution and memory access logic, and these units collectively form a pipeline. There are several compute cores residing within each SM, and each compute core is a large, heavily-pipelined execution unit capable of executing both integer and floating point operations. When a kernel is launched, each CTA is dispatched to an SM and executes there until its completion.
A CTA is further divided into units called warps, also known as wavefronts. Typically, there are a large number of warps in flight inside an SM so that memory access latency can be masked by concurrent execution. There is a unified L2 cache that is coupled with memory controllers, while the global memory is off-chip. On-chip data communication is implemented through an on-chip network.

\begin{figure}[t]
  \centering
  \scalebox{1.1}{\epsfig{file=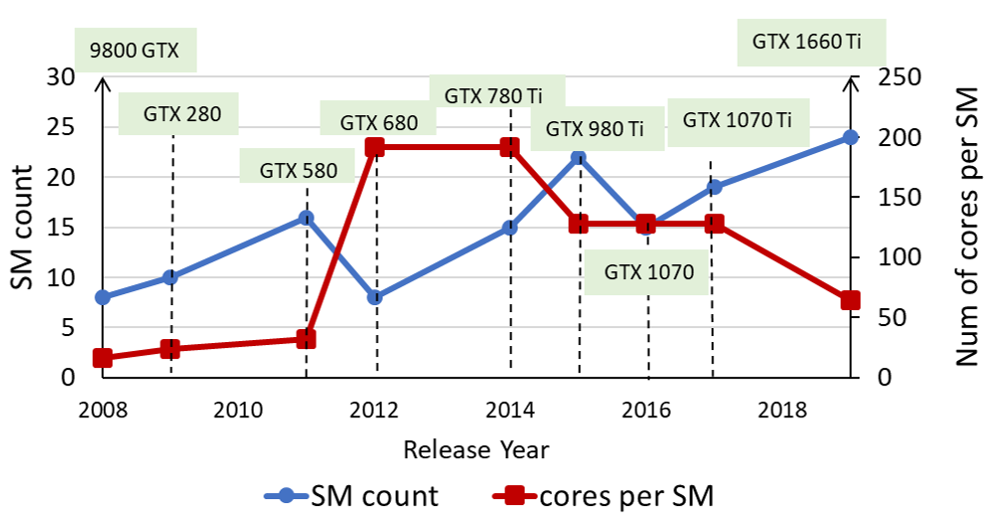,width=0.45\textwidth,clip=}}
   \vspace{-0.28in}
 \caption{ SM scaling trends (number of SMs vs numbers of cores per SM) for NVIDIA's GTX GPU family\cite{GTX}. }.
   \label{fig:GTX}
\vspace{-0.35 in}
\end{figure} 

\subsection{Scaling of SMs}
The primary execution unit in a GPU is a compute core, and they are grouped into SMs that share resources such as register file, local memory, and warp scheduler.
Due to the resource limitation, once the chip size is decided, the total number of compute cores is fixed. Then, there arises an important scaling problem: {\em how should compute cores and other resources be partitioned into SMs?} That is, should we opt for {\em scale-up} SMs (by including more compute cores and resources into a smaller number of SMs) or {\em scale-out} SMs (by having more SMs with fewer cores and less resource inside)? The scaling of configuration of SMs is critical since it directly determines the maximum parallelism among GPU threads and impacts resource sharing and utilization.

Figure~\ref{fig:GTX} shows the scaling trends of NVIDIA's GTX GPU family during the last 11 years. The number of computing cores per SM can be used to represent the scaling in SM size. We observe that both size and number of SMs are increased from 2008 to 2011. However, after 2011, the trends of SM size and SM count start to part their ways in opposite directions. This is because we are reaching the limit in terms of the total number of computing cores that can be integrated into a chip due to power and area constraints.  Therefore, it is not possible to scale up both the size and number of SMs; so, we either scale out or scale up, but not both,  as shown in the figure. And, the most recent trend has been scaling out since 2017. However, the question is whether this trend of scaling out is sustainable for the future. And, if not, what is the optimal configuration for the best performance and resource utilization?
\section{Motivation}
\subsection{Trade-offs in SM Scaling}
\begin{figure}[t]
  \centering
  \scalebox{1}{\epsfig{file=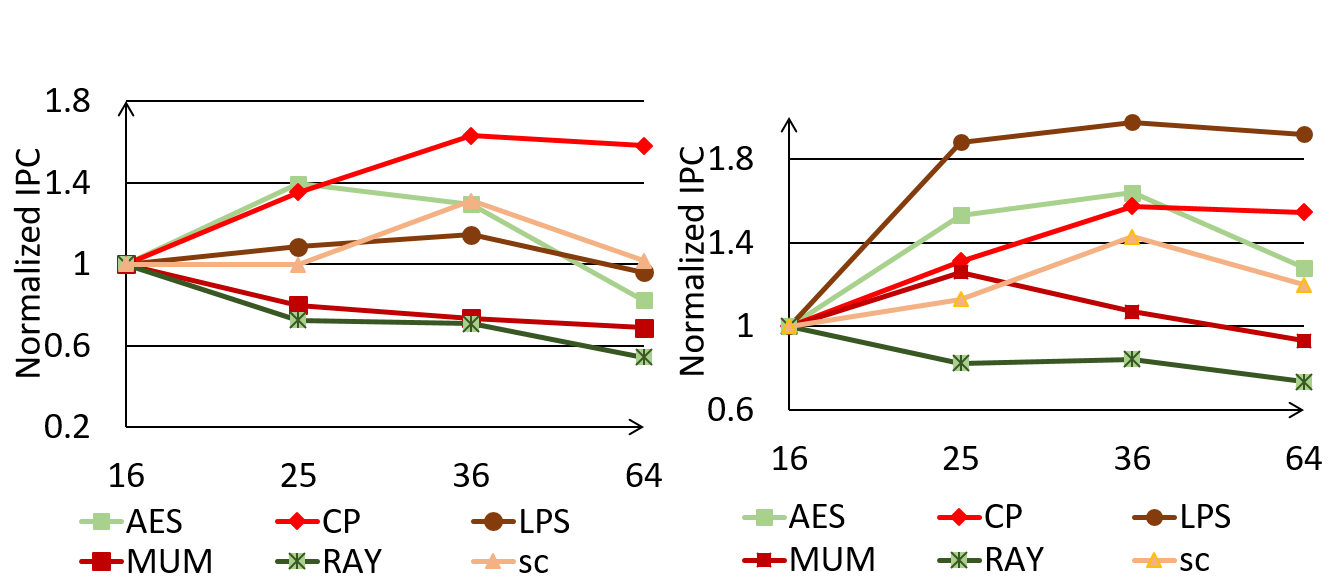,width=0.45\textwidth,clip=}}
   \vspace{-0.5cm}
 \caption{Performance with SM scaling (a) with a mesh NoC (b) with a perfect NoC. (x-axis is the number of SMs and y-axis is IPC normalized to 16 SMs.)} 
   \label{fig:scale}
\vspace{-0.2 in}
\end{figure}

As discussed above, warps execute in SMs and all threads in one SM share GPU resources such as shared memory, L1 cache, register file, warp scheduler, and interconnect interface. Scaling of SM greatly influences the resource utilization and power-performance efficiency. Due to their different characteristics and resource requirements, different applications exhibit {\em varying} SM scaling patterns.
We start by investigating the scaling of multiple benchmarks, and the results are plotted in Figure~\ref{fig:scale}(a). In this experiment, we \textit{fit the total amount of chip resources} but vary the size and the number of SMs. As can be observed, some applications benefit from scaling out with smaller SMs (\textit{CP, SC}), while other applications benefit from scale-up SMs (\textit{MUM, RAY}). This result indicates that there is not a scaling setting that benefits all applications. Motivated by this observation, we next examine in detail the major factors that determine an application's performance with SM scaling.

\begin{figure}[t]
  \centering
  \scalebox{1}{\epsfig{file=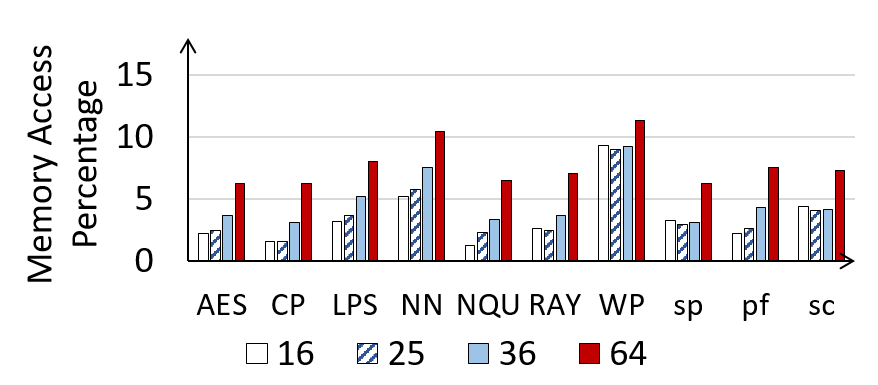,width=0.45\textwidth,clip=}}
   \vspace{-0.5cm}
 \caption{Memory access coalescing results with different GPU scaling options. Actual memory access rate represents the memory accesses \textit{after} coalescing. Here, we experiment with different SM scaling options with 16, 25, 36, and 64 SMs.} 
   \label{fig:coalescing}
\vspace{-0.2 in}
\end{figure}

\begin{figure}[t]
  \centering
  \scalebox{1}{\epsfig{file=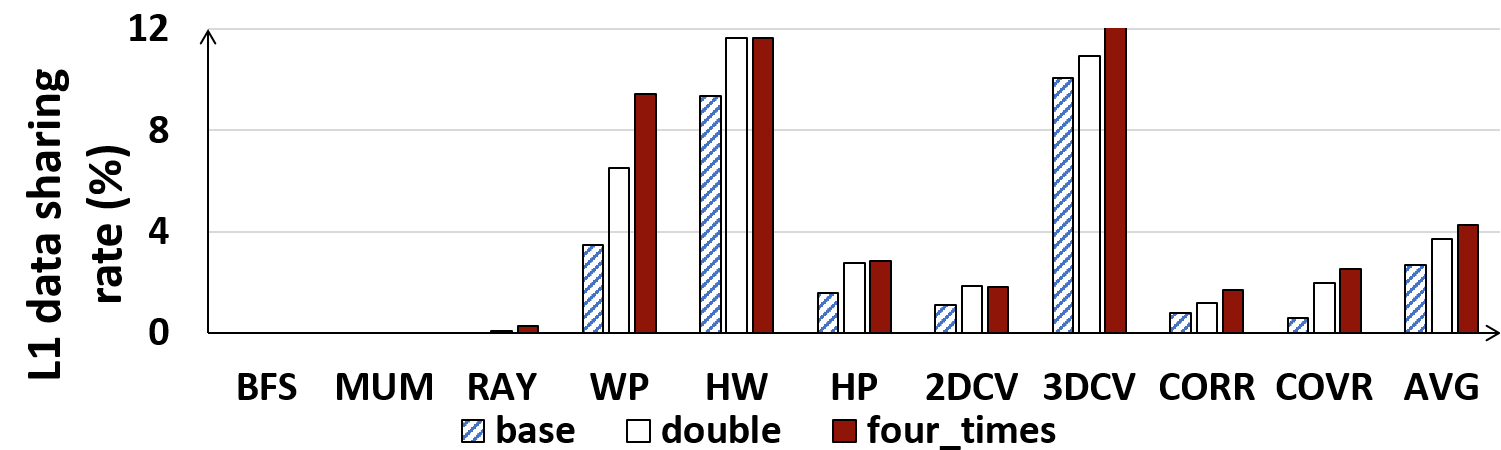,width=0.45\textwidth,clip=}}
   \vspace{-0.5cm}
  \caption{Rate of shared data in L1 caches of neighboring SMs.} 
   \label{fig:data_share}
\vspace{-0.15 in}
\end{figure}

\begin{figure}[t]
  \centering
  \scalebox{1}{\epsfig{file=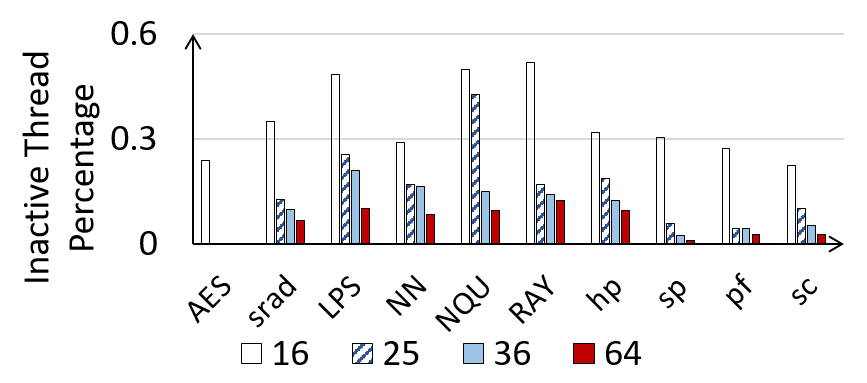,width=0.45\textwidth,clip=}}
   \vspace{-0.5cm}
 \caption{Control divergence caused stalls with different GPU scaling options.} 
   \label{fig:ctrl}
\vspace{-0.27 in}
\end{figure}

\textbf{(1) NoC Effect on SM Scaling.}     GPU SMs are connected to L2 cache and memory controllers through a network-on-chip (NoC). It has been shown that NoC is a bottleneck in GPU performance as the chip size grows \cite{checkerboard}. This is due to the particular traffic pattern exhibited by GPUs. Specifically, all SMs communicate with the limited number of memory controllers on chip. As the total on-chip bandwidth is fixed and is shared by all SMs, more SMs means that each SM receives a smaller share of the network bandwidth. In addition, a larger network incurs longer delays due to increased hop count and contention. As a result, there will be more negative impact on the performance. We experimented with different SM scaling options using a {\em perfect NoC} (with zero delay), and the results are plotted in Figure~\ref{fig:scale}(b). We can observe that when the NoC impact is removed, more applications (e.g., \textit{LPS, AES, CP,} and \textit{SC}) achieve better performance with scale out settings. This means that, for applications that are sensitive to the on-chip network performance, performance will ultimately degrade when we keep scaling out the SMs. 

\textbf{(2) Memory Locality and SM Scaling.}
It has been observed that memory resources inside an SM affect the performance of some applications. 
Some applications may share data a lot among warps in one SM or among L1 caches in different SMs. 
In such cases, scaling up will improve the utilization of shared memory and L1 cache, then reduce accesses to memory outside of an SM. GPUs employ a mechanism called \textit{memory coalescing} to reduce data movements. The idea is to combine multiple memory accesses from a warp to the same cache line into a single transaction. 
Larger SMs can execute larger warps, and provide more opportunities for memory coalescing. We quantitatively characterize the coalescing effects in SMs with different scaling settings, as shown in Figure~\ref{fig:coalescing}. In this figure, the y-axis shows actual memory access percentage of all load and store instructions after coalescing. As shown in Figure~\ref{fig:coalescing}, a scale up design with 16 SMs has much lower memory accesses compared to a scale out design with 64 SMs. That is, as far as coalescing is concerned, scale up SMs bring more benefits than scale out SMs.

In addition, recent GPU architectures combine data cache and shared memory functionality into a single memory block to provide the best overall performance. This makes the actual L1 cache capacity several times larger when needed. For example,
NVIDIA's Volta \cite{volta} architecture has a combined capacity of 128 KB/SM, more than seven times larger than the GP100 data cache, and all of it are usable as a cache
by programs that do not use shared memory. Considering this trend, we also investigated L1 data sharing among neighboring SMs with increased L1 capacity, and the results are plotted in Figure~\ref{fig:data_share}. As can be observed, some benchmarks (such as \textit{HW} and \textit{3DCV}) exhibit around 10\% sharing rate in the baseline configuration. When L1 capacity is increased by two or four times, higher sharing
rate is observed in most benchmarks that exhibit data sharing. This means that scaling up SMs by increasing the L1 capacity can effectively reduce duplicated data and
leads to more efficient utilization of the L1 caches.

\textbf{(3) Control Divergence and SM Scaling.}
Recent GPU architectures allow individual threads to follow distinct program paths with control flow on the SIMD pipeline. 
Control divergence occurs when threads in the same warp take different paths upon a conditional branch, which can lead to significant performance degradation because it increases pipeline stalls \cite{control1, control2}.
Even though various software techniques have been proposed to better schedule branch instructions \cite{software1, software2, software3}, control divergence cannot be totally removed. We have observed the core inactivity caused by the control divergence, as shown in Figure~\ref{fig:ctrl}. It can be seen from this plot that, for scale up SMs, pipeline stalls caused by branch instructions are much larger than scaling out SMs. In fact, for many benchmarks, the cores are stalled for more than half of the time waiting for branch instructions to be resolved. This is because, in larger SMs, the pipeline is wider than smaller SMs; as a result, a pipeline stall causes more reduction in computation parallelism. In this sense, applications with many control instructions need to employ scale out SMs for better performance.

\subsection{Accommodating Control and Memory Divergence through Dynamic SM Scaling}
Ideally, threads running on GPUs are able to execute in a lock step fashion,  and consume continuous computation enabled by warp scheduling to avoid pipeline stalls.
However, it has been shown that, for some applications, control flow divergence and memory divergence inside warps can significantly degrade performance by causing stalls in SM pipelines \cite{control1, control2, dynamicwarp1}. 
Memory divergence occurs when threads from a single warp experience different memory-reference latencies caused by cache misses or accessing different DRAM banks. In current organizations, the entire warp must wait until the last thread to have its reference satisfied.
To solve this problem, several techniques have been proposed to divide a warp into smaller slices and regroup them to create a new warp so that divergent threads do not prevent other threads from proceeding in execution \cite{warpslicer,GPUMaestro, DWS}. However, to our knowledge, all existing work subdivides a warp and reorganizes the threads to build a new warp to run on the "same sized" SM. 

\vspace{-0.1 cm}
There is a significant drawback of the above mentioned techniques when implementing variable warp sizes: the SM needs to be subdivided to support the execution of a gang of split warps. For example, the gang-warp \cite{GPUMaestro} needs to divide an SM into four slices and each slice works, after splitting, as a small SM. There are prohibitive hardware overhead and design complexity issues in this type of approach. In addition, prior work only considers resource utilization inside an SM, but not across SMs.
In contrast, we consider sharing among SMs at a larger granularity. We also consider resources such as NoC, L1 sharing, and coalescing among SMs, which have never been explored by prior work. 
\vspace{-0.1 cm}

In our proposed approach, we first observe the application's scalability with SM resources such as network and memory. If we detect that the application works better with scale up cores, we \textit{fuse} two small SMs into one big SM.
However, such a scheme fuses all SMs statically and may not flexibly adapt to a program's dynamic divergence. For example, some control and memory divergence between the threads inside a warp may cause long stalls in the fused SM since the pipeline is much wider now.  Based on this observation, we propose to dynamically \textit{split} the scaled-up SM into two smaller SMs to handle the control and memory divergence within a warp. Once we detect that the divergence no longer exists, we {\em fuse} the two SMs back to one. 
\vspace{-0.1 cm}

Since the fused SM already consists of two sets of execution paths, there is no extra hardware needed to support slicing, as opposed to the prior work \cite{warpslicer}. 
It needs to be noted that, we \textit{dynamically} split and fuse SMs independently in this scheme. Fusing and splitting decisions are made based on the current warp's running status, locally on each SM. As a result, when using our approach, at any given time during execution, the GPU architecture can have two types of SMs: some (fused) big SMs and some (split) SMs. Through this type of \textit{dynamic heterogeneity}, we are able to further improve resource utilization and achieve better performance, over state of the art.

\subsection{Reconfiguration Overhead} 
Usually, reconfigurable architectures involve redesigning micro-architecture units, and this may lead to significant overhead if not handled carefully. Due to this reason, there have not been many reconfigurable CPU architectures proposed in the past. However, in the case of GPUs, the reconfiguration overhead can be much lower.  This is because GPU SMs have much simpler structure and control logic, compared to general out-of-order CPU cores. Specifically, a GPU has a very simple in-order pipeline which reduces the reconfiguration complexity. In addition, GPUs are designed to hide memory latency by overlapping the execution of a large number of threads. As a result, delays caused by reconfiguration can be conveniently masked. This makes GPUs excellent candidates for reconfigurable architectures. Reconfiguration overhead also heavily depends on the granularity at which reconfiguration takes place. In this work, we propose a coarse-grained reconfigurable architecture based on SMs, which can further reduce design complexity and overhead. Specifically, we only reconfigure GPUs at an SM level without modifying pipeline structures. We only modify a few managed resources such as warp queues,  L1 cache, and register files. Therefore, the proposed GPU architecture is very amenable to reconfiguration.

%% file: design.tex
\section{Dynamic SM Scaling through Reconfiguration}

The goal of our design is to reduce resource under-utilization and also improve performance. To reduce the design complexity, we opt for coarse-grain reconfiguration.
Since it has been shown that individual kernels exhibit regular behavior, we propose a one-time reconfiguration scheme on a kernel-by-kernel basis. Once a kernel is determined to benefit from scale up SMs, we fuse every two \textit{neighboring} SMs to create scale up SMs. Otherwise, we continue executing the kernel using scale out SMs.
Our method is basically a top-down approach: we first characterize the kernel's overall scaling behavior regarding overall GPU resource utilization and then make a decision regarding whether to fuse or not. Based on this static fusion scheme, we also propose to refine the mechanism by allowing individual fused SMs to split dynamically if warps exhibit significant divergence in the fused SM. 


\subsection{Online Reconfiguration Controller}
\begin{figure}[t]
  \centering
  \scalebox{1}{\epsfig{file=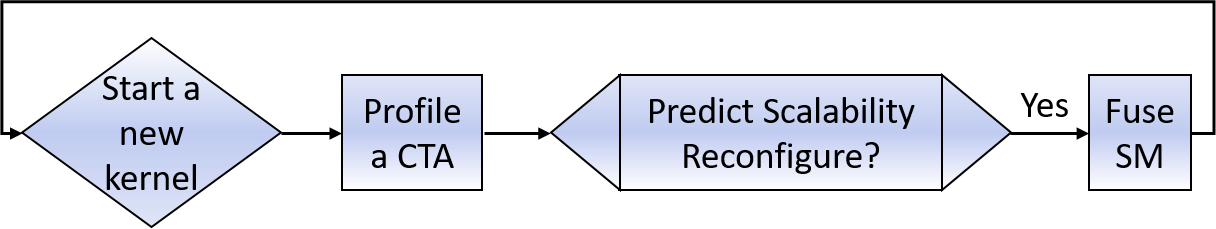,width=0.4\textwidth,clip=}}
   \vspace{-0.3cm}
 \caption{ Reconfiguration controller overview.} 
   \label{fig:flowchart}
\vspace{-0.18 in}
\end{figure}

A high level view of our reconfiguration controller is shown in Figure~\ref{fig:flowchart}. Profiling has been employed by many resource utilization techniques to determine an application's characteristics \cite{GPUMaestro, corefusion, compositecores}. In this work, we propose to combine online profiling with an offline trained model to predict scalability. When a new kernel starts, we first evaluate various metrics regarding its execution. Then, these metrics are fed into a scalability predictor which is already trained offline. The scalability predictor gives a result indicating whether the kernel should be executed on scale up or scale out SMs. Next, we reconfigure the SMs according to this result and start executing the kernel. After the kernel finishes, we start the loop again for the next kernel.
\subsubsection{Online Scalability Sampling}
\begin{figure}[t]
  \centering
  \scalebox{0.9}{\epsfig{file=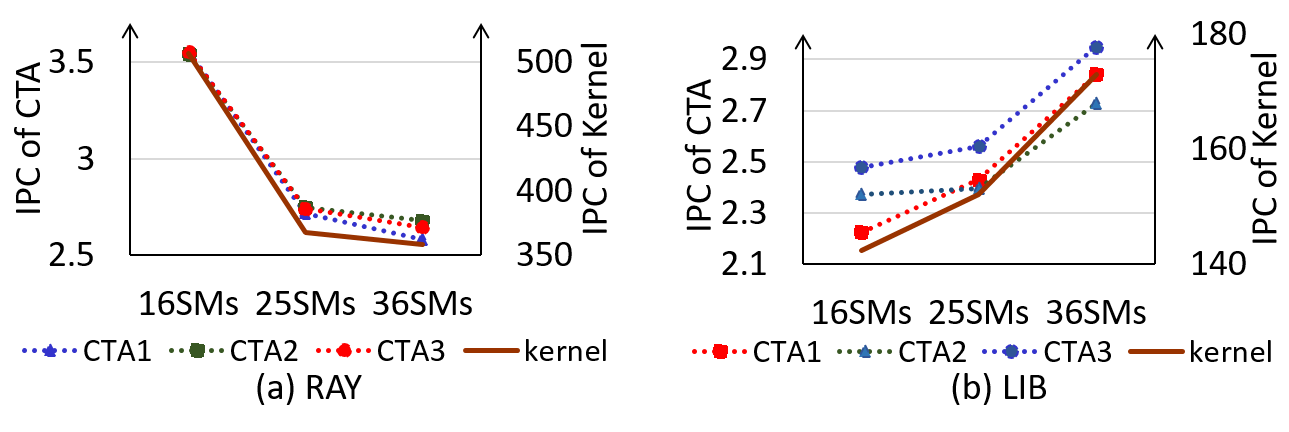,width=0.5\textwidth,clip=}}
   \vspace{-0.5cm}
 \caption{ Kernel and CTA scalability consistency.}
   \label{fig:cta-kernel}
\vspace{-0.25in}
\end{figure}

It has been shown that kernels exhibit disparate behavior with SM scalability and resource utilization \cite{SMK, gpu-cta}. Therefore, we cannot profile kernels to predict the behavior of an entire application. Recall however that, each kernel is split into smaller blocks, called CTAs, that execute similar portions of the code.
We found that the CTAs exhibit very consistent behavior, which closely follows the scalability trend at the kernel granularity.
Figure~\ref{fig:cta-kernel} shows how CTAs follow the same scaling trend with their kernel using applications \textit{LIB} and \textit{RAY}. As can be observed, both the kernel and CTAs of \textit{RAY} show a scale up trend, whereas \textit{LIB} kernel and its CTAs exhibit a scale out trend. Therefore, we propose to use a CTA to predict the scaling behavior of a kernel.

\subsubsection{Scalability Metrics}
To profile an application's scalability respect to the SM size and number, we need to identify metrics that can influence the scalability. Following are the major metrics we considered in this work:

\textcircled{1} NoC throughput:
This metric reflects the application's "communication intensity". If the NoC is a bottleneck, choosing scaled up cores will improve performance because the SM count would be smaller and the network size would accordingly be smaller, resulting in each core having a higher network throughput.


\begin{figure*}[t]
  \centering
  \scalebox{1}{\includegraphics[width=0.7\textwidth]{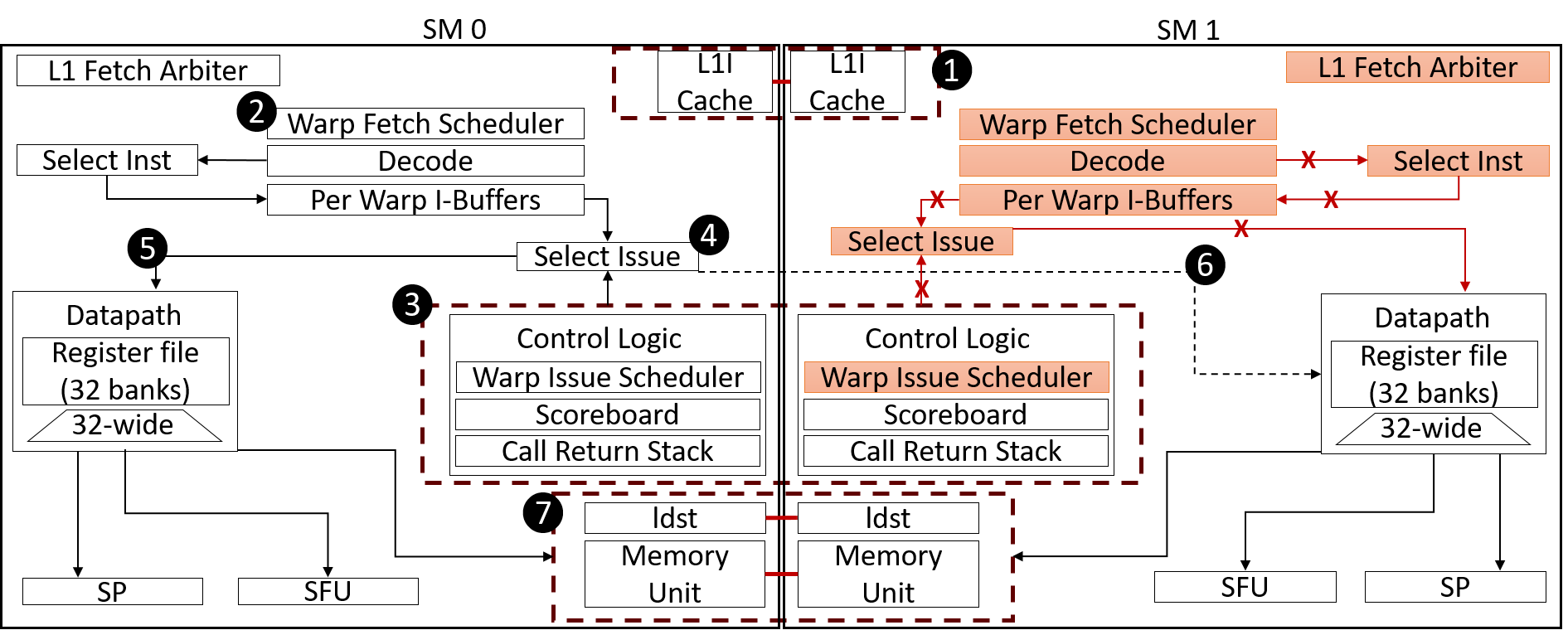}}
   \vspace{-0.3cm}
 \caption{SM reconfiguration via fusion.} 
   \label{fig:corefusing}
\vspace{-0.25 in}
\end{figure*}

\textcircled{2} Average NoC latency:
This is the average latency of the packets. It can also be used to evaluate the communication intensity.
\textcircled{3} Coalescing rate:
The coalescing rate is calculated as the number of actual memory accesses sent out from each SM divided by the total number of memory accesses in the instructions. This metric reflects how much shared data are requested across warps in an SM. 
\textcircled{4} L1 cache miss rate:
This reflects the demand for an application on local memory. If the miss rate is high and the data is not streaming, allocating a larger L1 will improve the performance, which means scale-up SMs are expected to have better performance.
\textcircled{5} MSHR rate:
This metric is similar to the coalescing rate, but it is across different instructions. Scale up SMs will have more instructions running on the fly, and this will benefit the  applications with higher MSHR rates.
\textcircled{6} Inactive thread rate:
This is used to reflect the warp control divergence. It is calculated as the number of cycles threads spent idling due to control instructions, divided by the total execution cycles. Kernels with larger control divergence would favor scale-out SMs.
\subsubsection{Scalability Predictor}
In this work, we propose to use \textit{binary logistic regression}, which is a machine learning technique borrowed from the field of statistics, to predict scalability. Our model accepts several input parameters and generates a binary output indicating whether an application needs to be run with scaled up GPUs or scaled out GPUs. 
Since we only fuse two neighboring SMs to build a scale up core, we only need a simple regression based model to predict scalability. The output of the model is only Binomial: yes or no to scale up.

Binary logistic regression estimates the probability that a characteristic is present (e.g., estimating the probability of "success"), given the values of explanatory variables. Unlike the normal distribution, the mean and variance of the Binomial distribution are {\em not} independent. Specifically, the mean is denoted by $P$ and the variance is denoted by $P*(1-P)/n$, where $n$ is the number of observations, and $P$ is the probability of the event occurring (i.e. whether we need to reconfigure smaller SMs into bigger SMs) in any one trial. If we were considering the data in a
list rather than a table form, we would assume that the variable had a mean $P$, and a variance $P*(1-P)$, and this variable would have a Bernoulli distribution.  When we have a proportion as a response, we use a logistic or logit transformation to link the dependent variable to the set of explanatory variables.
The logit link has the form:
\begin{equation}
    Logit (P) = \log [P/(1-P)].
\end{equation}
The term within the square brackets is the odds of an event occurring. In our case, it indicates whether we need to configure bigger cores. Using the logit scale changes the scale of a proportion to plus and minus infinity, and also because of Logit ($P$) = 0, when $P$ = 0.5. When we transform our results back from the logit (log odds) scale to the original probability scale, our predicted values will always be at least 0 and at most 1.
If there is only one input $x$, then we can write the model as:
\begin{equation}
  P=\frac{e^{(b_0 + b_1x)}}{1 + e^{(b_0 + b_1x)}},
\end{equation}
where $y$ is the predicted output, $b_0$ is the bias or intercept term, and $b_1$ is the coefficient for the single input value ($x$). 
We can write the model in terms of odds as:
\begin{equation}
    \frac{P}{1-P}= e^{b_0 + b_1x}.
\end{equation}



Conversely, the probability of the outcome \textit{not} occurring is 
\begin{equation}
1-P=\frac{1}{1+e^{b_0 + b_1x}}.
\end{equation}



For an event with multiple input factors, the modeled logarithm of the chance is given by:
\begin{equation}
    \log (\frac{P}{1-P}) = b_0 + b_1x_1 + b_2x_2 + ... +b_nx_n + constant, 
\end{equation}
where $P$ indicates the probability of an event (e.g., chance to scale up by fusing SMs in our case), and $P_i$ are the regression coefficients associated with the reference group and the $x_i$ explanatory variables.  We train this binary logistic model using a large amount of offline experimental data to obtain the values of $b_0-b_n$. We then use this model to directly infer the fusing decision online. Since the model is in fact linear, its implementation overhead is quite low. We give more details of the overheads in later sections.

\subsection{Design of the Reconfigurable Architecture}
The goal of AMOEBA is to create a GPU architecture that can dynamically change the number and size of its SMs, based on run-time workload behavior. We propose to start with a "baseline" scale out machine and fuse the neighboring SMs into a bigger SM, if the application is found to perform better with scaling out. Note that we allow fusing only two \textit{neighboring} SMs. This is due to the following considerations:
(1) Our scale out SM has 32 SIMD units and a scaled up SM will have 64 SIMD units when two SMs are fused. Fusing more SMs would significantly increase the pipeline width and the probability of pipeline stalls. In the future, if the scale out SM gets even smaller, for example, with 16 SIMD units, then fusing 4 such SMs together will be a more viable option. Note that our techniques can be easily extended to fusing more SMs to scale up. (2) Because fused SMs share resources such as L1 cache, register files, and warp schedulers, fusing more SMs means increased communication latency and implementation complexity. For example, a larger L1 cache will need a longer access time which will compromise the potential benefit from the SM fusion.
Due to these reasons, we only consider fusing two neighboring SMs in this paper.

Figure~\ref{fig:corefusing} shows how two scaled out SMs are fused to create a scaled up SM. The dashed lines show the fused units of the two SMs, placed to ensure that they can work in a lockstep fashion as one SM. The shaded components in SM1 are disabled due to SM fusion. In the fused SM, instructions are first fetched from the fused L1 I-cache (\ballnumber{1}). Then, the instructions are decoded, and selected instructions are sent to the per warp I-buffers (\ballnumber{2}). Next, the control logic (\ballnumber{3}) decides which instruction to issue and the decision is sent to the issue unit (\ballnumber{4}). Selected warps are then sent to the datapath of both SM0 (\ballnumber{5}) and SM1 (\ballnumber{6}) for execution. Memory accesses are sent from the executing threads to the fused memory unit (\ballnumber{7}).

In Figure~\ref{fig:corefusing}, there are two baseline SMs, shown as SM0 and SM1. 
AMOEBA does not change the execution units such as SP or SFU. When fused, the register files of the two original SMs and score boards work independently, as in the baseline.
AMOEBA does not change register files, and since the register files, are allocated with warps, they are not fused but can be accessed independently. Thus, there is no change in the throughput of any individual register file. Similarly, the score board connection with each register file is not modified either. However, the connection of the score board in SM1 to the warp scheduler is removed when two SMs are fused (\ballnumber{3}). Instead, this score board is connected to the warp scheduler of SM0.
This is because when we fuse two SMs, only one warp scheduler is kept, and it schedules all warps on both the SMs (\ballnumber{2}).

The memory components of the two SMs need to be fused, and this includes the shared memories, L1 I caches, L1 D caches, and L1 context cache.
We fuse L1 caches by increasing the cache associativity. To reduce the new L1 cache access latency, the SM layout needs to be modified as shown in Figure~\ref{fig:corefusing}, so that the L1 caches of both SMs are placed next to each other (\ballnumber{1}, \ballnumber{7}). Since the GPUs are good at hiding memory access latencies through overlapped warp execution, the extra delay caused by accessing a larger L1 D cache can be hidden by warp computation. In our experiments, we conservatively added one extra cycle in L1 cache access due to the cache fusion. Our results show that this extra delay is hidden quite well by the overlapped computation.

\begin{figure}[t]
  \centering
  \scalebox{1}{\epsfig{file=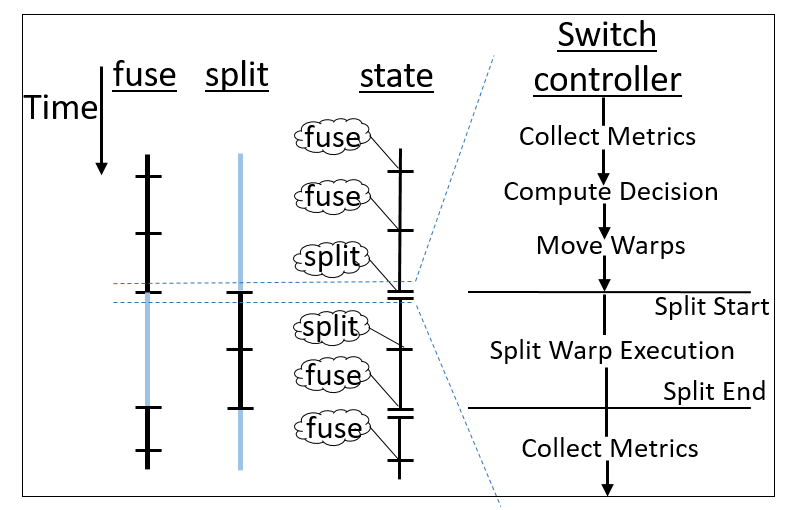,width=0.38\textwidth,clip=}}
   \vspace{-0.3cm}
 \caption{ Mechanism for switching between fusing and splitting.} 
   \label{fig:timing}
\vspace{-0.18 in}
\end{figure}

\begin{figure}[t]
  \centering
  \scalebox{1}{\epsfig{file=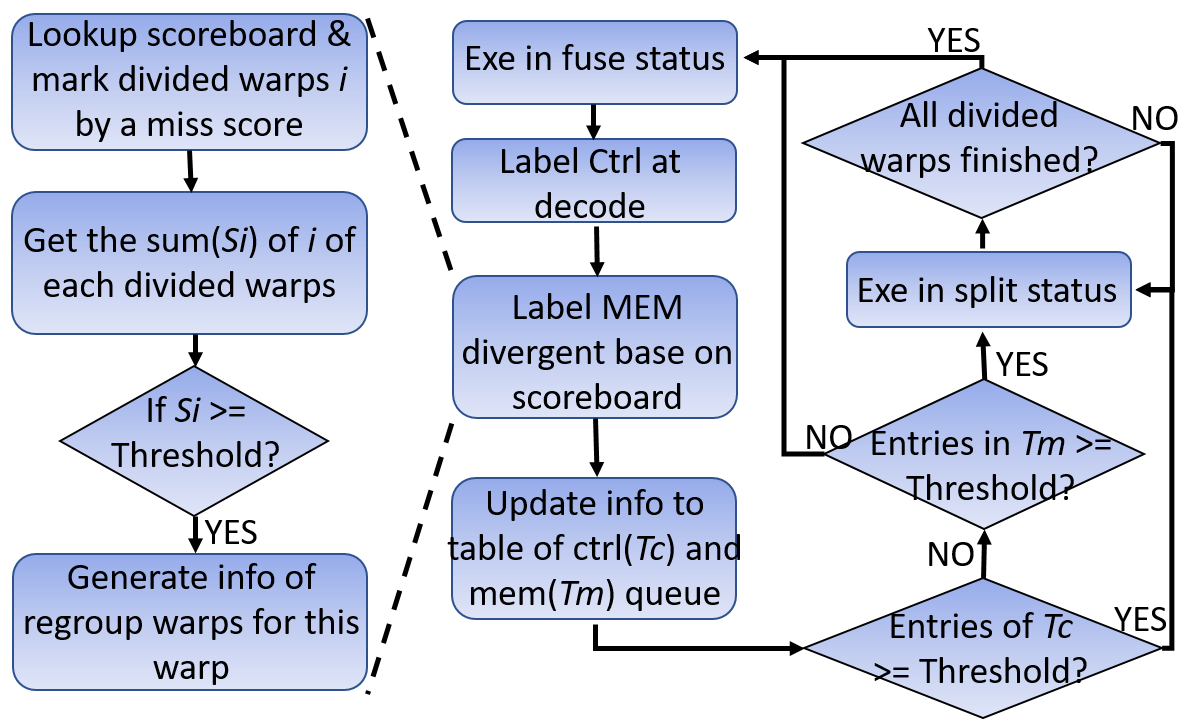,width=0.38\textwidth,clip=}}
   \vspace{-0.3cm}
 \caption{ Algorithm to dynamically split a fused SM to accommodate warp heterogeneity.} 
   \label{fig:algorithm}
\vspace{-0.25 in}
\end{figure}




Each fused SM has one copy of the coalescing unit in the fused core by fusing the two coalescing units from both the SMs. Since the warp size is doubled, this leads to more chances for coalesced memory accesses. After fusing the two SMs, AMOEBA combines the NoC routers of the two SMs into one by disabling one SM's router. This is implemented by adding a bypass path in one disabled router. As a result, the network size is reduced, this significantly reduces the network latency, and consequently, each router can enjoy a higher throughput in the network.

\subsection{Enabling Heterogeneous SMs through Dynamic SM Fusing and Splitting}
We propose to fuse SMs to reconfigure the GPU as a scaled up architecture when we observe that fusing the resources of two SMs is beneficial from a performance angle. It needs to be noted that our approach is different from prior works such as variable warps \cite{control1} or warp subdivision \cite{DWS}. Those works only consider the resources \textit{inside} an SM and try to fully utilize them -- there is \textit{no} cross-SM resource utilization optimization in those prior studies. Our proposed architecture, on the other, hand takes into account cross-SM resource utilization, such as NoC resource, sharing L1 caches between SMs, and memory access coalescing across SMs. As a result, it is fundamentally different from the earlier works.

However, there are still opportunities to further improve resource utilization in AMOEBA. This is, when we fuse two SMs, there can be scenarios where warp heterogeneity can cause inefficient pipeline utilization. For example, even though fusing two SMs can bring benefits in cache access or NoC, the resulting larger warp size creates wider pipelines. In this case, divergence in memory or control behavior in warps could lead to more pipeline stalls, compared to the unfused SMs. Therefore, we propose a dynamic SM splitting strategy: when we observe a significant warp divergence, and wide pipeline leads to a higher performance degradation compared to the benefits from fusion, we split the fused SM into two separate SMs. In this way, each split SM has half the pipeline width and the warps that cause divergence can only cause stalls in one of the smaller SMs. The other SM can keep the computation without being delayed by the pipeline stalls.


We can have different policies to decide when to split a given "fused" SM into two independent ones. Note that, by "independent", we mean that two SMs are running different warps independently on their respective data paths.
However, to reduce the cost of hardware and context switch, we do not split the shared resources, such as L1 cache, register files, and NoC interface.
We set up a threshold to decide when to split, which is a fixed ratio of divergent warps to the total warps running in the large SM. 
If the current ratio is greater than the threshold, we decide to split the SM into two. This figure also shows how NoC interfaces are bypassed when two SMs are fused together.

After the SM splits, we move all divergent warps from the bin to a new SM created from the split.
Subsequently, the two SMs start the independent execution of their warps.
When the second SM finishes all divergent warps,
we \textit{re-fuse} the two SMs into one.
Then, we start the procedure to collect divergent warps again and split the SMs when necessary.
Thus, this procedure of splitting and fusing is dynamically decided by the divergence of warps. 
This mechanism is expected to maximize resource utilization and reduce stalls in the fused SMs.

\begin{table}[b]
\vspace{-0.3in}
\centering \scriptsize
\caption{\label{table:setup} System configuration. See GPGPU-Sim v3.2.2 \cite{gtx-480} for the full list.}
\begin{tabular}{|p{0.22\textwidth}||p{.2\textwidth}|} \hline
{Number of Computing Cores} &  48 cores\\
\hline
{Number of Memory Controllers} &  8\\
\hline
{MSHR per Core} &  64\\
\hline
{Warp Size} &  32\\
\hline
{SIMD Pipeline Width} &  8\\
\hline
{Number of Threads per Core} &  1024\\
\hline
{Number of CTAs/Core} &  8\\
\hline
{Constant Cache Size/Core} &  8KB\\
\hline
{Texture Cache Size/Core} &  8KB\\
\hline
{L1 Cache Size/Core} &  16KB\\
\hline
{L2 Cache Size/Core} &  128KB\\
\hline
{Number of Registers/Core} &  16384\\
\hline
{Warp Scheduler} &  Greedy-Then-Oldest\\
\hline
{Shared Memory} &  48 KB\\
\hline
{Memory Scheduler} &  FR-FCFS\\
\hline
{Memory Model} &  8 MCs, 924 MHz\\
\hline
{NoC Channel Width}& 128 bit\\
\hline
{NoC Topology}& mesh\\
\hline
{NoC Router Pipeline Stage}& 2\\
\hline
\end{tabular}
\vspace{-0.2in}
\end{table}

The idea behind splitting is to prevent divergent warps from causing pipeline stalls. So, we need to separate divergent warps and non-divergent warps into two clusters and run each cluster on a separate smaller SM, so that the slow warps do not cause the fast warps to delay. Suppose that we have split a scale up SM into 2 scale out SMs (SM\_0 and SM\_1), and then, we want to run fast warps on SM\_0 and slow warps on SM\_1. 
There can be different mechanisms that can be used to decide what warps to be moved to the second SM\_1. In this work, we investigated two methods: (1) \textbf{direct split}, and (2) \textbf{warp regrouping}. The direct split method is simple as it directly divides a divergent warp in the middle into 2 smaller warps. Then, both the smaller warps are moved to SM\_1. This method has a low cost but may not have optimal performance. This is because the slow threads in a divergent warp may be located in different positions. If we simply cut the warp in half, there can be varying combinations of resulting warps. For example, we can have one warp with all fast threads and one warp with all slow threads. Or, we can have both smaller warps with partially slow threads. The ideal case is the first splitting, since we can better remove negative effects of the slow threads on the fast ones. Based on this analysis, we propose a second method that regroups threads into a fast warp and a slow warp. 
We then move the slow warp to SM\_1 and keep the fast warp in SM\_0.
To accomplish this, we first divide the threads in the original warp into small groups, and label them as "fast" or "slow" based on divergence. Then, we regroup them into two warps so that the slowest groups are all put into a slow warp and moved to SM\_1.
In our design, we also periodically check the stalls in the slow SMs.
We periodically move some fast warps to them so that the resources are not wasted when the slow warps cause stalls. 

The hardware overhead of the splitting is low because the split SMs were anyway two independent SMs in the baseline architecture. We added hardware to fuse them as described earlier, and splitting them does {\em not} need extra hardware, except the management and storage of the divergent warps. Therefore, we need a new warp queue and some simple control logic. Compared to the prior works \cite{control1, DWS} that proposed splitting resources inside one SM, our overhead is very low.
Figure~\ref{fig:timing} and Figure~\ref{fig:algorithm} show the timing and algorithm of our dynamic splitting and fusing.




%% file: evaluation.tex

\begin{figure}[t]
  \centering
  \scalebox{1.1}{\epsfig{file=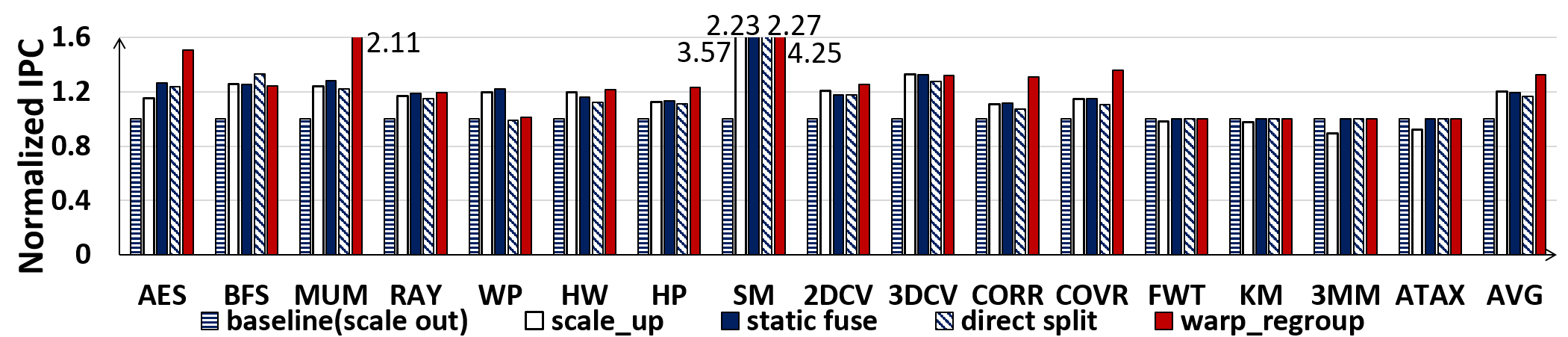,width=0.45\textwidth,clip=}}
   \vspace{-0.3 in}
 \caption{Performance results.}
   \label{fig:ipc}
\vspace{-0.18 in}
\end{figure}

\begin{figure}[t]
  \centering
  \scalebox{1.1}{\epsfig{file=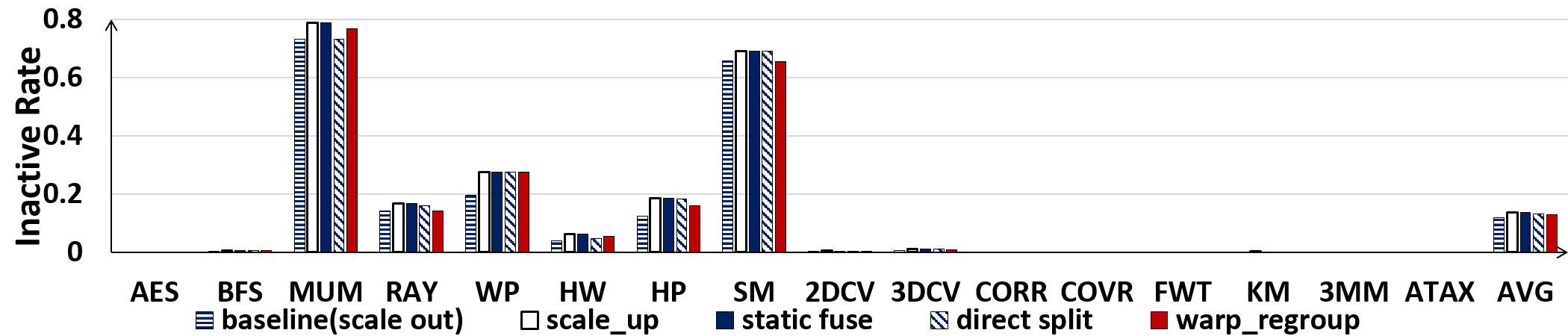,width=0.45\textwidth,clip=}}
   \vspace{-0.3in}
 \caption{Control divergence caused stalls.}
   \label{fig:ctrl-stall-perf}
\vspace{-0.25 in}
\end{figure}


\section{Experimental Evaluation}
We simulate our baseline architecture using a cycle-level simulator (GPGPU-Sim \cite{gpusimv}) and faithfully model all key parameters (Table 1).
The baseline GPU consists of 48 scale out SMs with a warp size of 32. There are 8 memory controllers on the chip.
The interconnection network is a mesh-based NoC. There are two subnets to avoid deadlock between request and reply messages. The router has a pipeline with 2 stages.
When we perform reconfiguration, two baseline SMs are fused to create one scale up SM.
We use a wide range of GPU applications from Ispass \cite{ispass}, Rodinia \cite{rodinia}, Polybench \cite{poly} and Mars \cite{mars}, to evaluate our design, and execute all applications to completion. We report performance results using the geometric mean of IPC speedup (over the baseline GPU). We also report other evaluation metrics provided by the simulator such as L1 cache miss rate, NoC latency, network injection rate, and SM idle rate.

\subsection{Performance}
\subsubsection{Performance Impact}
Figure~\ref{fig:ipc} illustrates the performance gains when using AMOEBA. The baseline is a scale out architecture and we also experiment with direct scale\_up.
We present the performance of applying three techniques proposed by AMOEBA: static fuse configures the SMs only once before a kernel's execution. Using the prediction model, AMOEBA predicts the scalability of application with SMs, and chooses to fuse two SMs or not. The next two techniques are based on the dynamic heterogeneous SM scaling. Direct split simply divides a divergent warp into smaller ones in the middle, whereas warp regrouping employs more complicated techniques to re-organize threads into a fast warp and a slow warp.
As can be observed, \textit{SM} achieves the highest improvement in performance, by 4.25 times. \textit{MUM} also achieves a significant performance improvement of 2.11 times. On average, all 12 benchmarks have around 47\% increase in IPC. 

Static fuse achieves almost same performance as direct scale\_up when larger SMs can bring performance benefits. For applications that can benefit from larger SMs, static fuse achieves almost the same performance gain as direct scale\_up. However, some benchmarks prefer scale out configurations, such as \textit{3MM} and \textit{ATAX}. Our fusing techniques all perform better than direct scale\_up (about 10\%) for these workloads. This shows that AMOEBA can accurately predict the applications' scalability and the correct reconfiguration can lead to performance gains. Some workloads are not sensitive to scaling such as \textit{FWT} and \textit{KM} and all AMOEBA techniques perform similar to the baseline. In general, direct split and static fuse bring similar benefits (on average) for most workloads, except \textit{BFS} and \textit{SM}. Some workloads such as \textit{WP} even experience performance degradation, which is mainly due to the fusion overhead. This is because this technique cannot dynamically react to workload behavior changes. On the contrary, warp\_regrouping achieves 16\% performance gain than direct split because it can accurately capture a workload's dynamic behavior caused by divergence. 

%
\vspace{-0.1 cm}
\subsubsection{Control Stall Impact}
Figure~\ref{fig:ctrl-stall-perf} plots the SM inactive rate caused by control divergence which is defined as the fraction of cycles that SMs are stalled due to control instructions. We can observe that only part of the workloads suffer from stalls caused by control divergence. For workloads that have control divergence caused stalls, dynamic fusion perform better than direct scale\_up and static fusing because they can dynamically adjust to the changes in control divergence. Warp\_regrouping performs better in more cases than direct split because fast and slow warps are allocated to different SMs. Among all cases, the baseline scale out configuration has the least amount of stalls because its pipeline width is always smaller than the other configurations.

\begin{figure}[t]
  \centering
  \scalebox{1.1}{\epsfig{file=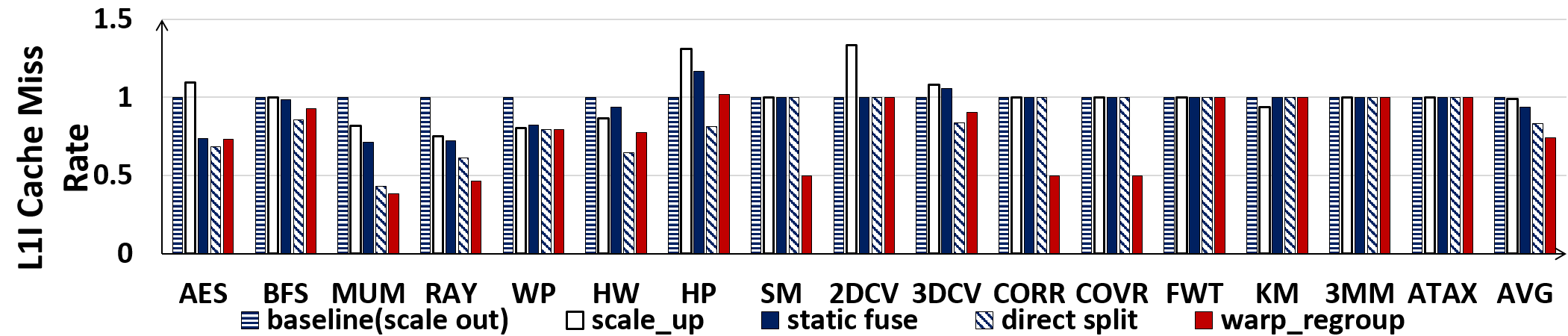,width=0.45\textwidth,clip=}}
     \vspace{-0.3 in}
 \caption{L1-I cache miss rate.}
   \label{fig:L1I-perf}
\vspace{-0.15 in}
\end{figure}

\begin{figure}[t]
  \centering
  \scalebox{1.1}{\epsfig{file=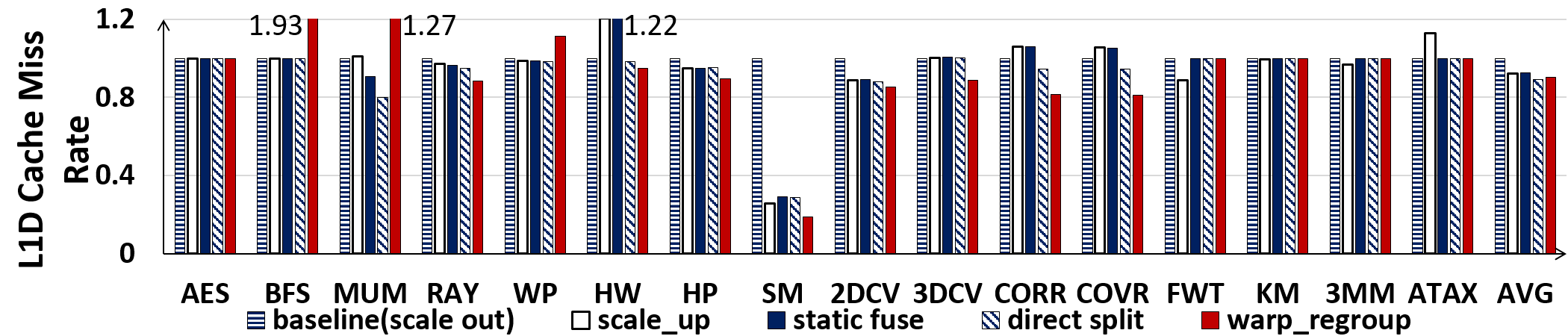,width=0.45\textwidth,clip=}}
     \vspace{-0.3 in}
 \caption{L1-D cache miss rate.}
   \label{fig:L1D-perf}
\vspace{-0.25 in}
\end{figure}


\begin{figure}[t]
  \centering
  \scalebox{1.1}{\epsfig{file=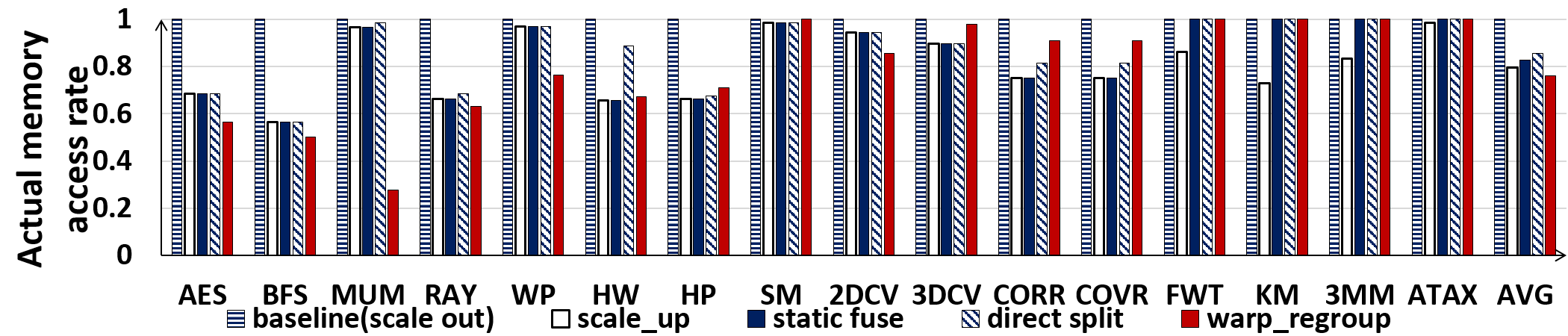,width=0.45\textwidth,clip=}}
   \vspace{-0.3in}
 \caption{Actual memory access.}
   \label{fig:mem-access-perf}
\vspace{-0.18 in}
\end{figure}

\begin{figure}[t]
  \centering
  \scalebox{1.1}{\epsfig{file=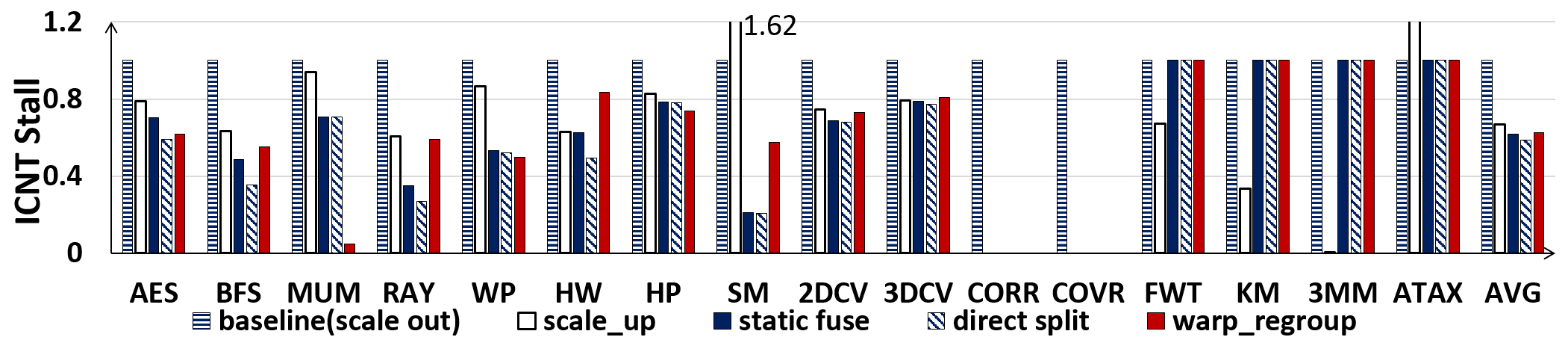,width=0.45\textwidth,clip=}}
   \vspace{-0.3in}
 \caption{Normalized rate of stalls when MCs cannot inject to the NoC.}
   \label{fig:noc-stall}
\vspace{-0.2 in}
\end{figure}

\subsubsection{Memory Access Impact}
L1-I cache miss rate is plotted in Figure~\ref{fig:L1I-perf}. Some benchmarks such as \textit{FWT}, \textit{3MM}, and \textit{ATAX} are not sensitive to L1-I cache capacity and fusing does not lead to any change in their behavior. However, most benchmarks have their miss rates reduced and the average reduction is 9\%, 20\% and 30\% for the three AMOEBA schemes.
Sharing L1-I cache through SM fusion reduces the I cache misses and thus leads to improved performance.
 Figure~\ref{fig:L1D-perf} plots the miss rate of L1-D cache.
The most significant reduction is for \textit{SM} and its miss rate is reduced by more than 70\%. This is because the sharing of L1 cache increases its effective capacity and this change directly leads to 4.25 times improvement in performance.
Some benchmarks, such as \textit{BFS} and \textit{MUM}, experience increased L1-D cache miss rates. This is because warp regrouping changes data locality by moving warps between SMs and this leads to higher miss rates.
Impact of AMOEBA on memory accesses is plotted in Figure~\ref{fig:mem-access-perf}. As can be observed, all benchmarks achieve reduced actual memory access rates compared to the baseline. Actual memory access rate is calculated as the actual memory access count divided by the total number of memory accesses in the instructions.
Since AMOEBA allows SMs to share coalescing units, the actual number of loads and stores is greatly reduced.
\vspace{-0.1 cm}
\subsubsection{NoC Impact}
\begin{figure}[t]
  \centering
  \scalebox{1.1}{\epsfig{file=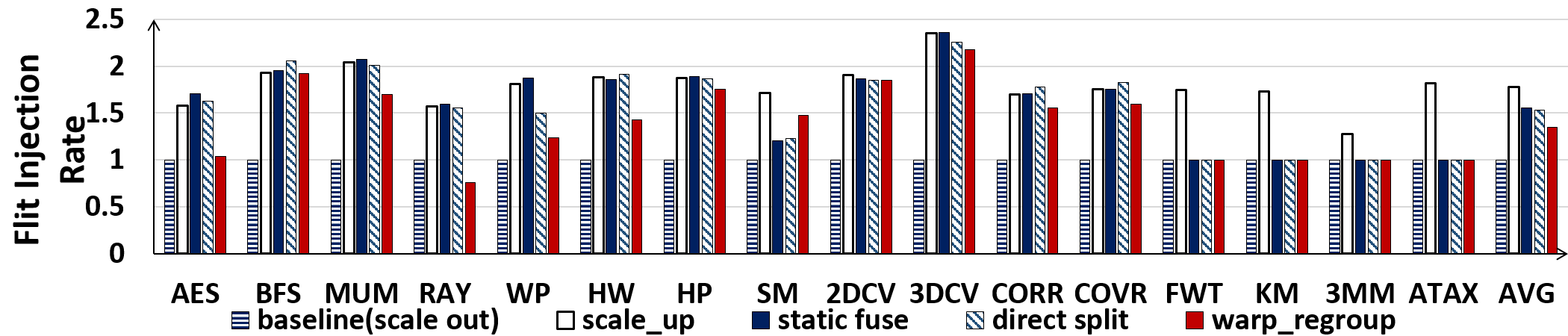,width=0.45\textwidth,clip=}}
   \vspace{-0.3in}
 \caption{NoC injection rate.}
   \label{fig:inject-perf}
\vspace{-0.21in}
\end{figure}

Figure~\ref{fig:noc-stall} plots the normalized ICNT stall rate, which is defined as the rate of stalls when new reply packets cannot be generated because an MC's injection queues are full. This data can reflect the pressure on both NoC and memory controllers.  As can be observed from this figure, all AMOEBA schemes are able to reduce this stall rate. For some benchmarks, such as \textit{CORR} and \textit{COVR}, this stall time is totally removed. Since AMOEBA can fuse SMs and bypass some routers, the network size is reduced, and this leads to smaller hop counts. As a result, NoC bottleneck can be greatly relieved for communication-intensive applications. Figure~\ref{fig:inject-perf} shows the average network data injection rates for the SM configurations evaluated.
As can be observed from this plot, all benchmarks have a higher injection rate under the AMOEBA than the baseline. This is because we fuse SMs and use only one NoC network interface to inject packets. Even though the injection rate is higher in AMOEBA schemes, the network size is reduced by half and this leads to shorter communication delays, paving the way to achieve better performance. 

\subsection{Dynamics of Core Fusion and Splitting}

\begin{figure}[t]
  \centering
  \scalebox{1}{\epsfig{file=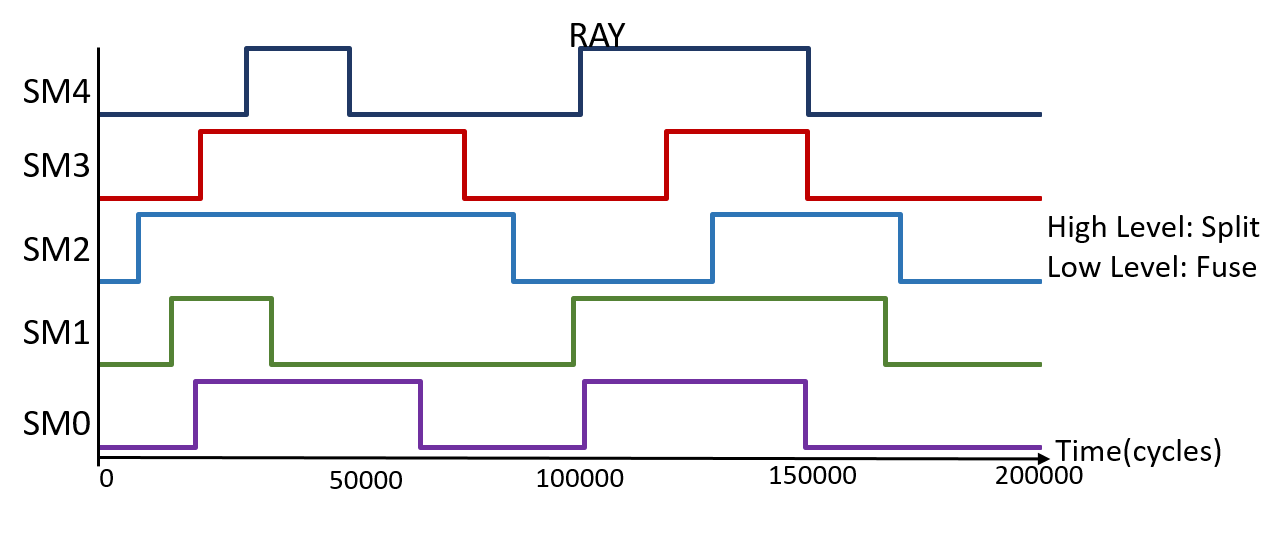,width=0.45\textwidth,clip=}}
   \vspace{-0.2in}
 \caption{Phases of dynamic SM fusion and splitting.}
   \label{fig:phase}
\vspace{-0.2in}
\end{figure}
To observe the dynamics of switching between fusing and splitting, we studied the status of five SMs in benchmark \textit{RAY}.
The results are shown in Figure~\ref{fig:phase}.
As shown in this figure, all 5 SMs start with fused execution because this benchmark favor scale up SMs.
After a period of time, the SMs start to split because enough divergent warps have been detected and smaller SMs brings more benefit.
However, the switching between fusing and splitting of each SM is independent of each other. As a result, at a certain time, there exist both scale up and scale out SMs in the architecture. As a result, better performance results are achieved from this flexible heterogeneity in SM configurations provided by AMOEBA.
\subsection{Analysis of Scalability Prediction Model }

\begin{table}[t]
\centering \scriptsize
\caption{\label{table:setup} Coefficients in scalability prediction model.}
\begin{tabular}{|p{0.12\textwidth}||p{.06\textwidth}||p{0.12\textwidth}||p{.06\textwidth}|} \hline
{Constant} &  -73.635 & {Concurrent cta} & 1.414\\
\hline
{Control Divergent} &  444.628 & {Coalescing} & 2057.050\\
\hline
{L1D Miss Rate} &  -313.838 & {L1I Miss Rate} & 1674.513\\
\hline
{L1C Miss Rate} &  -67.277 & {MSHR} & -102.971\\
\hline
{Load Inst Rate} &  -680.786 & {Store Inst rate} & -804.7\\
\hline
{NoC} &  -8.301 & & \\
\hline
\end{tabular}
\vspace{-0.1in}
\end{table}
We use several performance counters to generate the detailed metrics required by our scalability prediction model. Most of these performance counters are already included in many of today's GPU systems, including cache hits and misses, MSHR, and branch instruction statistics. For metrics cannot currently be provided by the performance counters, we propose to add such counters, e.g., concurrent CTA numbers. 
Table 2 shows the coefficients in our scalaiblity prediction model.

\begin{figure}[t]
  \centering
  \scalebox{0.8}{\epsfig{file=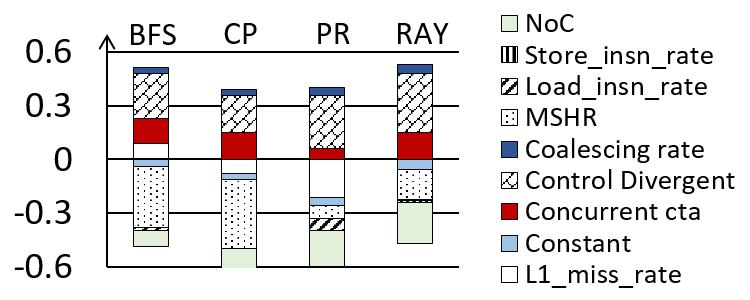,width=0.45\textwidth,clip=}}
   \vspace{-0.2in}
 \caption{Magnitude of parameter impact on determining scalability for some applications using the proposed predictor.} 
   \label{fig:magnitude}
\vspace{-0.18 in}
\end{figure}

To analyze the relative contribution of each metric to overall performance in the prediction model, we plot the distributed weights of the major metrics. Here, we consolidate different types of L1 cache miss rate into one metric called \textit{L1\_miss\_rate}.
The result is shown in Figure~\ref{fig:magnitude}.
For each metric, its magnitude of impact is shown as a value between -1 to 1. The magnitude of impact of a metric is calculated as $\textit{the coefficient of this }\\\textit{metric} \times \textit{measured value}$.
For example, the impact magnitude of Load instruction is calculated as $Load\_insn\_rate \\\times \textit{its coefficient}$. All positive impact magnitudes contribute to a scaling up decision, and all negative impact magnitudes contribute to a scaling out decision.
Eventually we add all metrics' impact magnitudes together and check the sum. If the result is positive, then we predict to fuse SMs and create a scale up configuration. Otherwise, we predict that a scale out configuration will fit better with the application.
In this figure, the sum of the impact magnitudes for \textit{BFS} and \textit{RAY} are both positive. So, these benchmarks favor running on scale up SMs. On the contrary, \textit{CP} and \textit{PR} prefer to run on scale out SMs.
It can also be observed that different applications' scalability is influenced by different metrics with varying extent. For example, MSHR plays a more significant role in \textit{BFS} and \textit{CP}, whereas \textit{PR} and \textit{RAY} are more sensitive to the NoC performance compared to others. 

\subsection{Comparing with State of the Art}

\begin{figure}[t]
  \centering
  \scalebox{1}{\epsfig{file=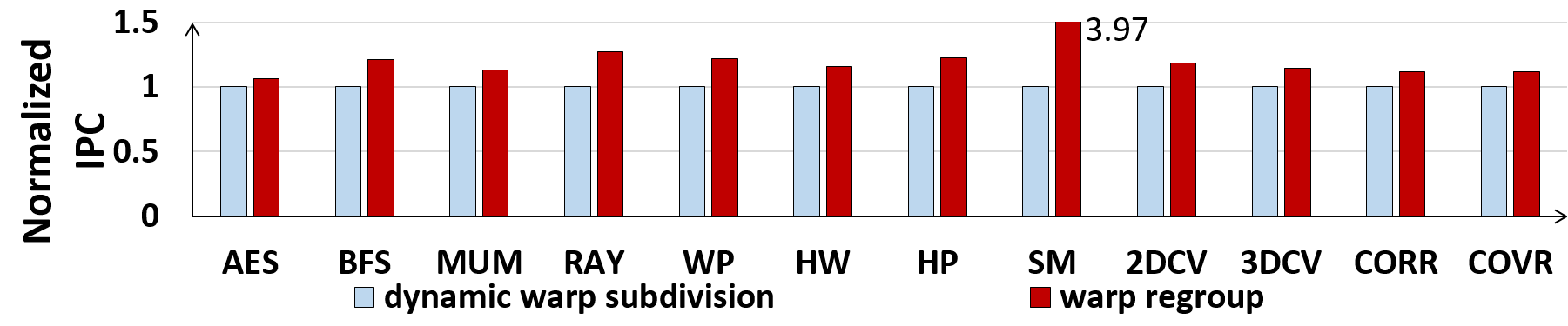,width=0.45\textwidth,clip=}}
   \vspace{-0.2in}
 \caption{Comparison with Dynamic Warp Subdivision (DWS) \cite{DWS}.}
   \label{fig:art}
\vspace{-0.28in}
\end{figure}
We now compare the performance of AMOEBA against Dynamic Warp Subdivision (DWS) \cite{DWS}. The results are plotted in Figure~\ref{fig:art}.
DWS was proposed by Meng et. al. to divide warps into smaller ones in order to reduce the stalls caused by memory and control divergence.
On average, AMOEBA achieves 27\% performance gain over DWS.
Benchmark \textit{SM} achieves 3.97 times improvement in performance compared to DWS. 
This is because DWS can only improve resource utilization \textit{inside} an SM and cannot harness the benefits of cross-SM resource sharing. 
In contrast, AMOEBA can dynamically change the configurations of SMs and thus flexibly allow resources to be shared among SMs. Thus, performance can be further improved through enhanced resource utilization.

\subsection{Area Overheads}
There are two types of controllers in the proposed architecture: online reconfiguration controller for scale up or scale out, and switch controller for dynamic fusing and splitting. We propose to implement these controllers in an IP module in the GPU chip.
The major components in the controllers are a MAC unit, buffers and control logic.
 We employ similar methods proposed in \cite{dynamicwarp1} to model the buffers in the controllers by using the area of a latch cell from the NanGate 45 nm Open Cell library.
The resulting area of each bit of the buffer is 4.2 um\textsuperscript{2}, 
and the total estimated added buffer area is 0.021 mm\textsuperscript{2}.
We use a pipelined Booth Wallas MAC \cite{MAC} and it is synthesized by Synopsis Design
Compiler using 90 nm technology and scaled to 45 nm. The area of the MAC is 0.019 mm\textsuperscript{2}. Together with the control logic, we estimate the two controllers to have area of 1.53 mm\textsuperscript{2}.
GeForce 8800GTX which has 128 SM cores, the overall area overhead of AMOEBA can be calculated as the total SM area overhead + controller overhead = 0.021 mm\textsuperscript{2} $\times$ 128 + 1.52 mm\textsuperscript{2} = 4.208 mm\textsuperscript{2}.
Compared to the total GeForce 8800GTX area of 480 mm\textsuperscript{2}, AMOEBA incurs an area overhead of 0.88\%.

%% file: relatedwork.tex
 \section{Related Work}
 There has been plenty of work proposing reconfigurable architectures for multi-core CPU systems \cite{voltron, trips, smartmemory, corefusion, composable}.
 A multicore architecture is proposed in \cite{voltron} that reconfigures cores into a wide VLIW machine to exploit hybrid forms of parallelism. 
 As a pioneer reconfigurable architecture, TRIPS \cite{trips} splits ultra-large cores to small ones to meet the diverse demand of application parallelism. 
 Working in the opposite direction of reconfiguring cores, Ipek et al. \cite{corefusion} proposed Core Fusion where a large core can be dynamically configured from a group of independent smaller cores. 
 Core Fusion is the most closely related work to AMOEBA, but it is proposed for CPU cores and the core fusing policy and micro architecture are very different from our work.

Compared to CPU based multicore systems, there have been fewer works on reconfigurable GPU architectures. 
Voitsechov et al. proposed SGMF, a dataflow architecture using coarse-grain reconfigurale fabric, composed of a grid of interconnected functional units \cite{SGMF}. However, SGMF needs help from compiler to break the CUDA/OpenCL kernels into dataflow graphs and integrates the control flow of the original kernel to produce a control-data-flow-graph (CDFG).
Different from their work, our proposed scheme does not require compiler support.
R-GPU is a reconfigurable GPU architecture that aims to reduce the cycles spent on data movement and control instructions and focus on data-computations \cite{rgpu}. It configures GPU cores to create a spatial computing architecture. 
R-GPU implements reconfiguration at a \textit{core level} within an SM, and does not consider an application's scalability while our work reconfigures at an \textit{SM level}, and our reconfiguration decision is based on NoC, control instructions and memory access patterns. 
Dhar et al. proposed fine grained and coarse grained reconfigurations of SMs in GPUs in order to reduce the underutilization of resources and power consumption \cite{datapathfusion}. However, their work only reconfigures the datapath inside each SM. Our work also reconfigures memory and NoC of the system, and we also propose to use heterogeneous SMs to improve performance and power efficiency. 

 Heterogeneous multicores have emerged as a promising approach for CPU-based systems which leverage cores with different capabilities and complexities to strike a balance between performance and power \cite{compositecores, Hill, kumar, biglittle, annavaram, balakrishnan, asymmetric}. 
 Lukefahr et al. propose composite cores that consist of big and small compute engines \cite{compositecores}. 
 Kumar et al. \cite{kumar} proposed a heterogeneous multi-core architecture to reduce power dissipation. 
 Hill et al. showed that there is great potential in performance improvement of the serial sections of an application using heterogeneous cores \cite{Hill}.
Our proposed AMOEBA architecture differs from these heterogeneous architectures in that our heterogeneous cores are dynamically configurable while these earlier works employ fixed core configurations.
Our design can provide more flexibility in exploring heterogeneous architectures, and achieve better resource utilization.

Recently, several approaches have been proposed for improving GPU resource utilization \cite{SMK, GPUMaestro, multiprogram, memorystall, isca_1, isca_2}.
Wang et al. propose Simultaneous Multikernel (SMK) by exploiting heterogeneity of different kernels \cite{SMK}. 
Park et al. proposed GPU Maestro that performs dynamic resource allocation for efficient utilization of multitasking GPUs \cite{GPUMaestro}. 
Wang et al. propose an application-aware TLP management techniques for a multi-application execution environment in order to make judicious use of shared resources \cite{multiprogram}. 
To improve resource utilization in concurrent kernel execution (CKE), Dai et al. proposed mechanisms to reduce memory stalls \cite{memorystall}.
Our proposed work is different from these prior techniques because it reconfigures SMs so that they scale according to the application's dynamic behaviour.

\vspace{0.1cm}
\section{Conclusion}
In this work, we propose a reconfigurable GPU architecture, called  AMOEBA, to explore the design space of GPU scaling. By predicting a given application's scalability with SM size, the proposed architecture is able to dynamically configure scale up or scale out SMs in order to achieve high performance and resource utilization.
We also propose an optimization strategy to further reconfigure each SM based on the warp divergence at run- time, resulting in a heterogeneous architecture in which \textit{both} scale up and scale out SMs co-exist at run-time.
Our evaluation results using various benchmark programs demonstrate the effectiveness of AMOEBA in reducing GPU resource under-utilization and improving system performance and power efficiency. 